\documentclass[]{spie}  
\usepackage[]{graphicx}
\usepackage{epstopdf}
\usepackage{amssymb}

\title{Teaching the concept of convolution and correlation using Fourier 
transform}

\author{Debesh Choudhury   
\skiplinehalf
 Adamas University, Department of Electronics and Communication Engineering\\ 
 Barasat-Barrackpore Road, PO - Jagannathpur\\ Pin - 700126, West Bengal, India
}

\authorinfo{Further author information: E-mail: 
debesh@iitbombay.org, Tel: +91-9831369809}

\begin{document} 
  \maketitle 

 \begin{abstract} 
 Convolution operation is indispensable in studying analog optical and 
digital signal processing. Equally important is the correlation 
operation. The time domain community often teaches convolution and 
correlation only with one dimensional time signals. That does not 
clearly demonstrate the effect of convolution and correlation between 
two signals. Instead if we consider two dimensional spatial signals, the 
convolution and correlation operations can be very clearly explained. In 
this paper, we propose a lecture demonstration of convolution and 
correlation between two spatial signals using the Fourier transform 
tool. Both simulation and optical experiments are possible using a 
variety of object transparencies. The demonstration experiments help to 
clearly explain the similarity and the difference between convolution 
and correlation operations. This method of teaching using simulation and 
hands-on experiments can stimulate the curiosity of the students. The 
feedback of the students, in my class teaching, has been quite 
encouraging.
 \end{abstract}


\section{Motivation}

In the under graduate programs of electrical, electronics and 
communication engineering, convolution and correlation are taught in 
many courses, such as signals and systems, digital signal processing and 
communication theory. The electronics or the time domain community treat 
the subject of convolution and correlation mainly with respect to 
one-dimensional time domain signals. On the other hand, the optics and 
photonics community teach convolution and correlation with respect to 
two-dimensional spatial signals. It is observed that the subject can be 
better appreciated if examples from two-dimensional spatial signals are 
considered.

\section{Introduction}
 \label{sec:background}

Convolution is a mathematical method of combining two signals to form a 
third signal. The characteristics of a linear system is completely 
specified by the impulse response of the system and the mathematics of 
convolution~\cite{DSP_book_Smith}. It is well-known that the output of a 
linear time (or space) invariant system can be expressed as a 
convolution between the input signal and the system impulse response 
function.

Convolution is the basis for many signal processing techniques. As for 
example digital filters are synthesized by designing appropriate impulse 
response functions. Targets are detected on the radar by analyzing the 
measured impulse responses. In long distance telephone communication, 
echo suppression is achieved by creating impulse responses those cancel 
out the impulse responses of the reverberation signal.

Correlation is a simple mathematical operation to compare two signals. 
Correlation is also a convolution operation between two signals. But 
there is a basic difference. Correlation of two signals is the 
convolution between one signal with the functional inverse version of 
the other signal. The resultant signal is called the cross-correlation 
of the two input signals. The amplitude of cross-correlation signal is a 
measure of how much the received signal resembles the target signal. The 
correlation peak specifies the location of the target. 

Correlation operation is regulalrly done in radar 
communication~\cite{DSP_book_Smith}. Vander Lugt evolved with a 
novel way of optical character recognition by optical matched filtering 
what he called ``complex spatial filtering''~\cite{VanderLugt}. Goodman 
and Weaver demonstrated how to optically convolve two spatial signals by 
joint Fourier transformation~\cite{Goodman_JFT}. The joint Fourier 
transform method forms the basis for optical implementation of 
cross-correlation between two signals, which is popularly known as 
joint-transfrom correlation~\cite{Goodman_book}.

It is worth noting that both convolution and correlation operation can 
be realized by applying Fourier transform. Thus, it is appropriate to 
explain the similarity and difference of convolution and correlation 
using Fourier transform. In this paper, we present a teaching method for 
understanding the concept of convolution and correlation using the 
Fourier transform tool.

\section{Mathematical expressions of convolution and correlation}

The mathematical expression defining a convolution between two continuous 
time signals $x(t)$ and $h(t)$ is given by~\cite{Goodman_book}
 \begin{eqnarray}
 y(t) &=& x(t) \otimes h(t) \nonumber \\
     &=&\int^{+\infty}_{-\infty} x(\tau) h(t-\tau) d\tau
 \end{eqnarray}

\noindent where $\otimes$ represents convolution operation. For discrete 
time signals $x[t]$ and $h[t]$, it can be expressed as a convolution sum 
given by
 \begin{equation}
 y[n] = \sum^{k=+\infty}_{k=-\infty} x[k]h[n-k]
 \end{equation} 

\noindent Equations~(1) and (2) also give the output of a linear time 
invariant system where $h(t)$ / $h[t]$ is the system impulse response of 
the continuous / discrete linear time invariant system.
 \begin{figure}[htb]
 \centering
 \includegraphics[width=5cm]{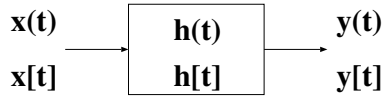}
 \caption{Linear time invariant system.}
 \end{figure}

\noindent The correlation between the same two signals may be expressed 
as
 \begin{eqnarray}
 c(t) &=& x(t) \ast h(t) \nonumber \\
     &=&\int^{+\infty}_{-\infty} x(\tau) h(\tau - t) d\tau \nonumber \\
     &=& x(t) \otimes h(-t)
 \end{eqnarray}

\noindent where $\ast$ represents correlation operation. For discrete 
time signals $x[t]$ and $h[t]$, it can be expressed as~\cite{DSP_book_Smith}
 \begin{equation}
 c[n] = \sum^{k=+\infty}_{k=-\infty} x[k]h[k-n]
 \end{equation} 

\noindent Convolution and correlation are similar mathematical 
operations. Correlation is also a convolution operation between the two 
signals but one of the signals is the functional inverse. So, in 
correlation process one of the signals is rotated by 180 degree. This is 
the basic difference between convolution and correlation. It is 
interesting to note that convolution and correlation can produce 
identical results if the signals are rotationally symmetric.

\section{2D examples of convolution}

The time domain community treats it mostly with 1D signals. It is 
observed that the concept can't be appreciated easily using 1D signals. 
If we take examples of 2D signals, we can show the results pretty simple 
and the concept is easily understandable by the students. We use the 
convolution theorem of Fourier transform. This states that the Fourier 
transform of a product of two signals is the convolution of the 
respective Fourier transforms. The vice versa is also true.

\begin{figure}[htb]
 \centering
 \includegraphics[width=1.6in]{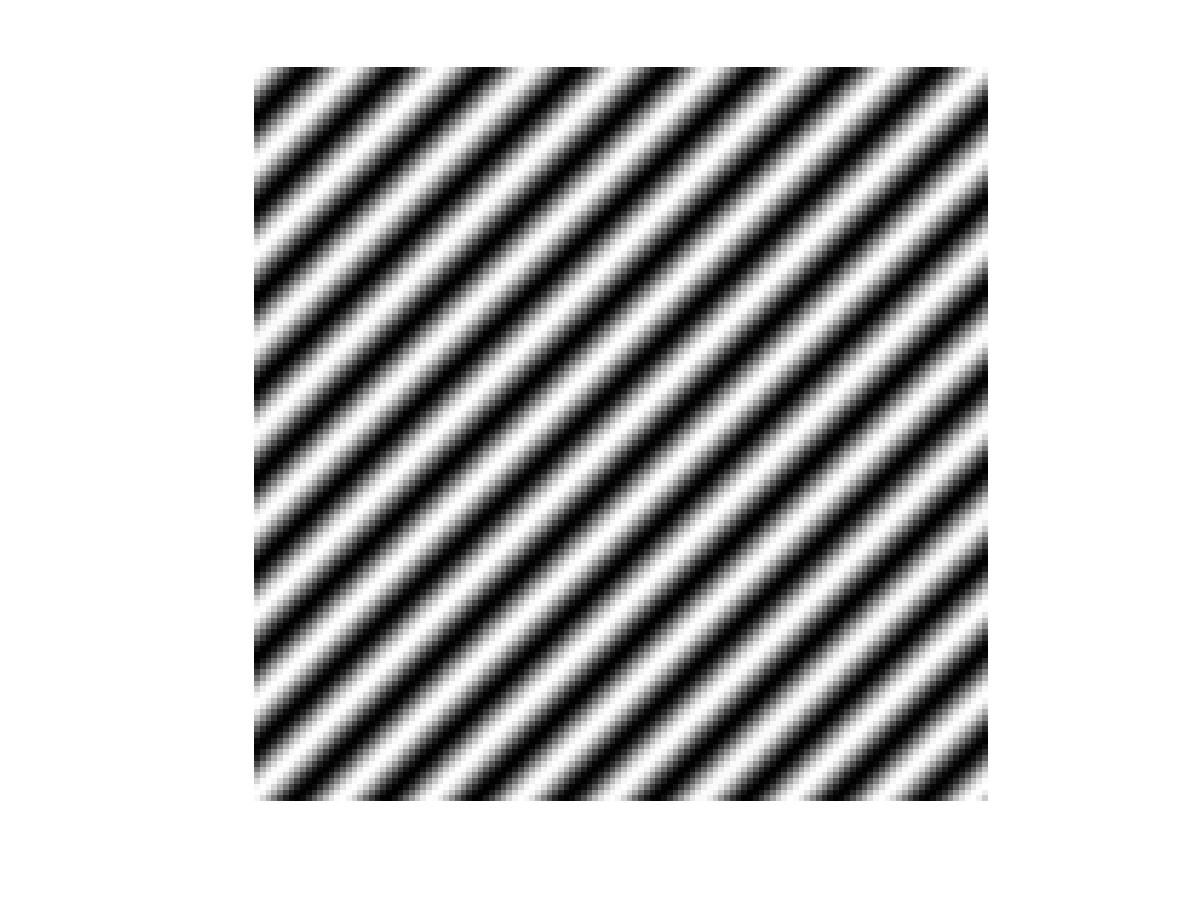}
 \includegraphics[width=1.6in]{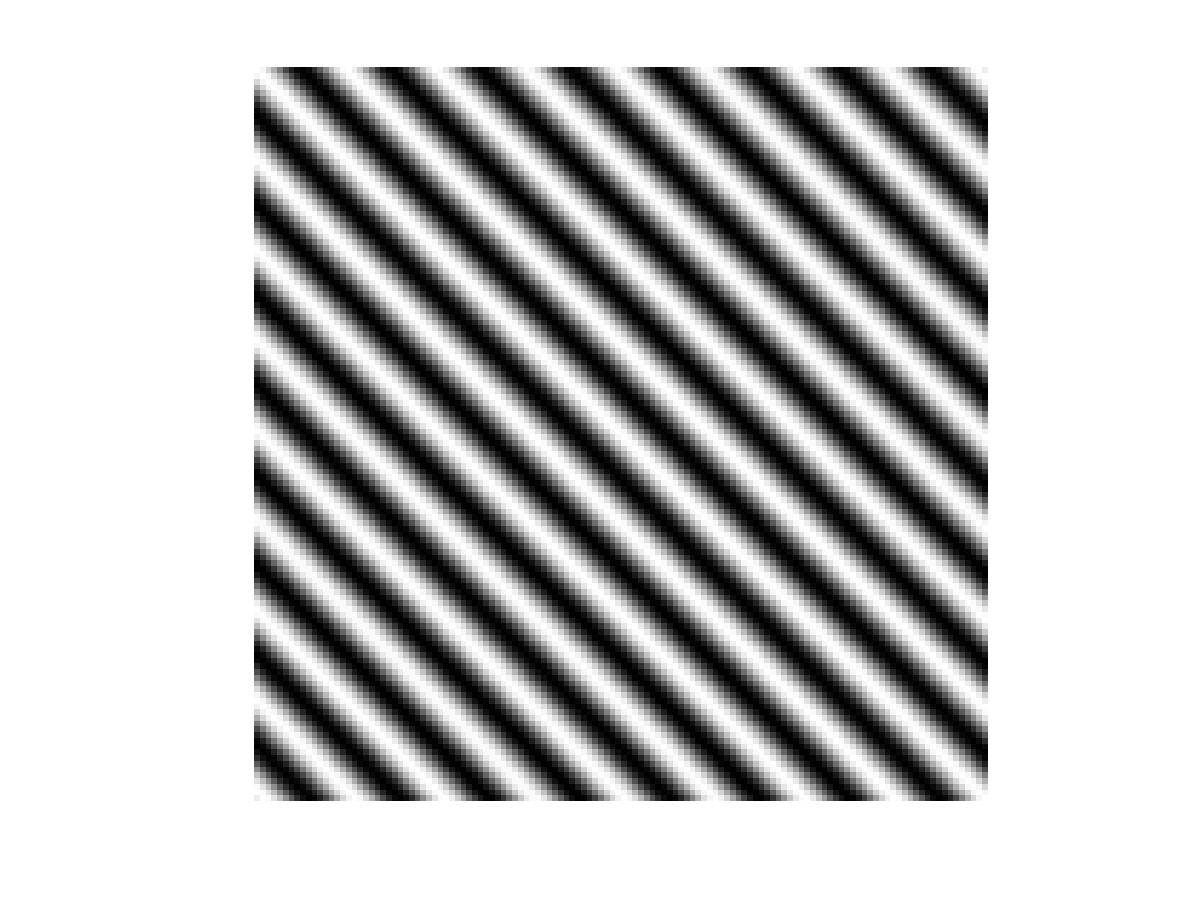}
 \includegraphics[width=1.6in]{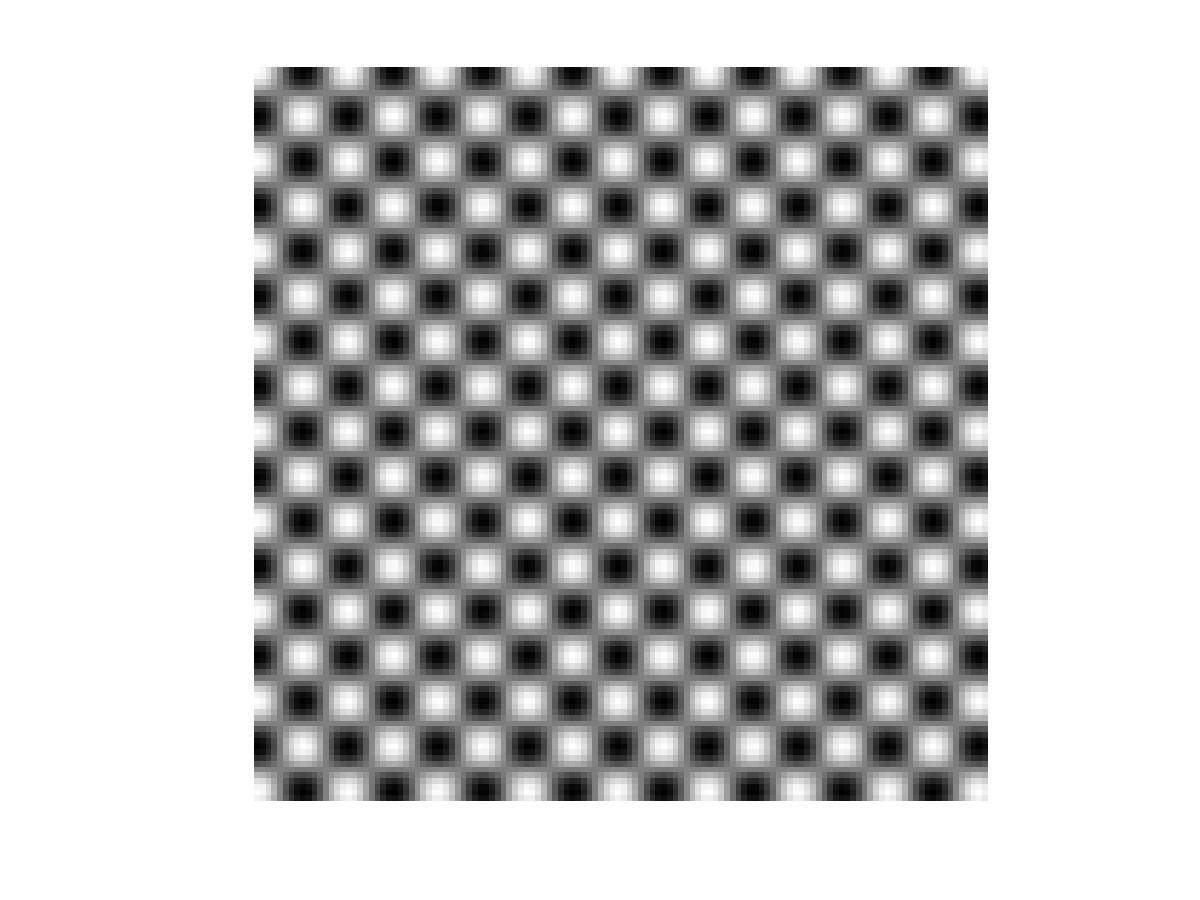}  
 \includegraphics[width=1.6in]{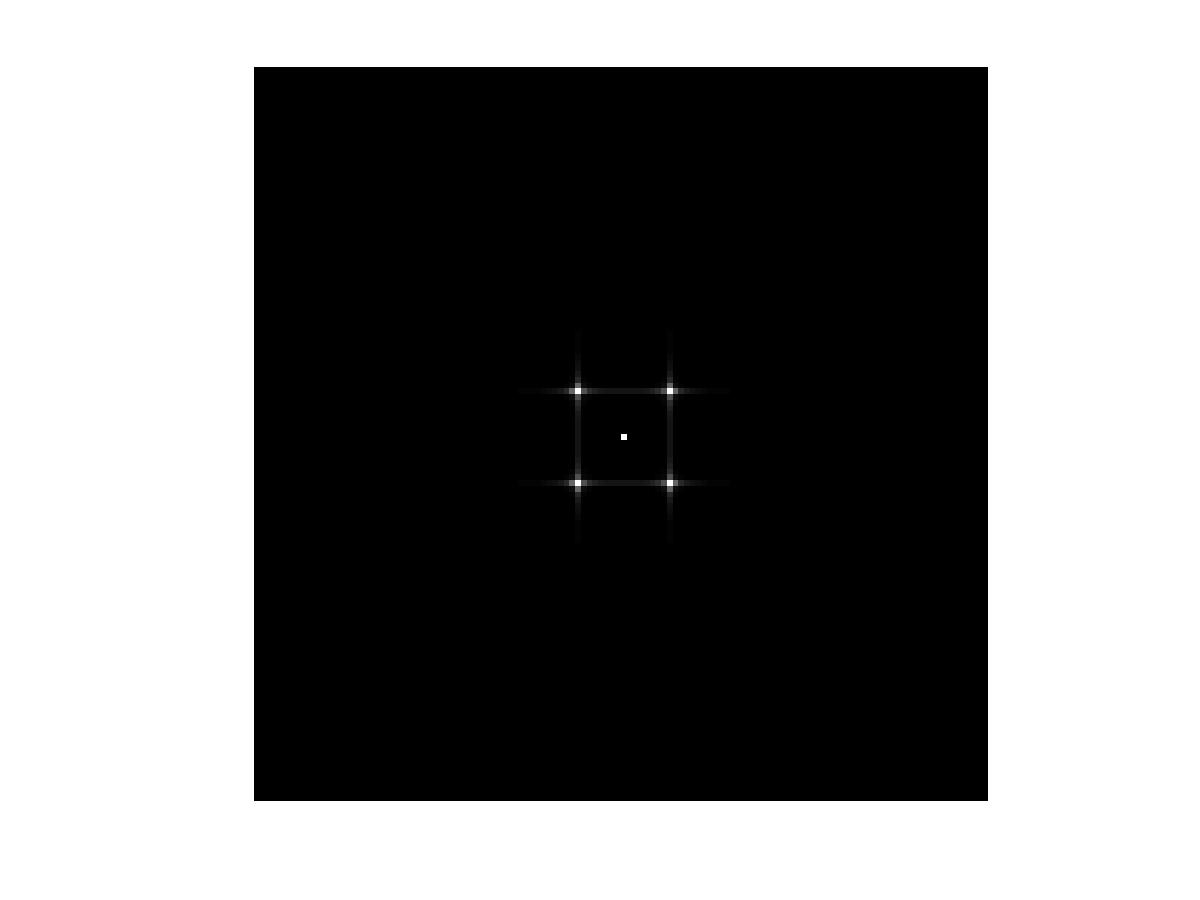} \\[-10pt]
 (a) \hglue 1.5in (b) \hglue 1.4in (c) \hglue 1.5in (d)
 \caption{Image of (a)~first grating; (b)~second grating; (c)~addition 
of (a) and (b); (d) { $ \vert {\rm FT} \vert $} of (c).}  \label{fig1}
 \end{figure}

Let us take two sinusoidal 2D gratings. First we add the gratings and 
take Fourier transform. If $g_1(x,y)$ and $g_2(x,y)$ be the amplitude 
transmittance of the two gratings, then we can write
 \begin{equation}
 {\mathcal FT} [ g_1(x,y) + g_2(x,y) ] = {\mathcal FT}[g_1(x,y)] + 
{\mathcal FT}[g_2(x,y)] \label{eq_add}
 \end{equation}

\noindent where ${\mathcal FT}$ is a Fourier transform operator. We 
created 2D sinusoidal gratings, the algebraic addition of the gratings 
and their Fourier transforms using open source GNU Octave~\cite{Octave}. 
The results are shown in Fig.\ref{fig1}.
 It is evident from Fig.\ref{fig1} that the Fourier transform of the 
addition signal is just the addition of the Fourier transforms of the 
individual gratings.

\begin{figure}[htb]
 \centering
 \includegraphics[width=1.6in]{./Pics/GX.jpg}
 \includegraphics[width=1.6in]{./Pics/GY.jpg}
 \includegraphics[width=1.6in]{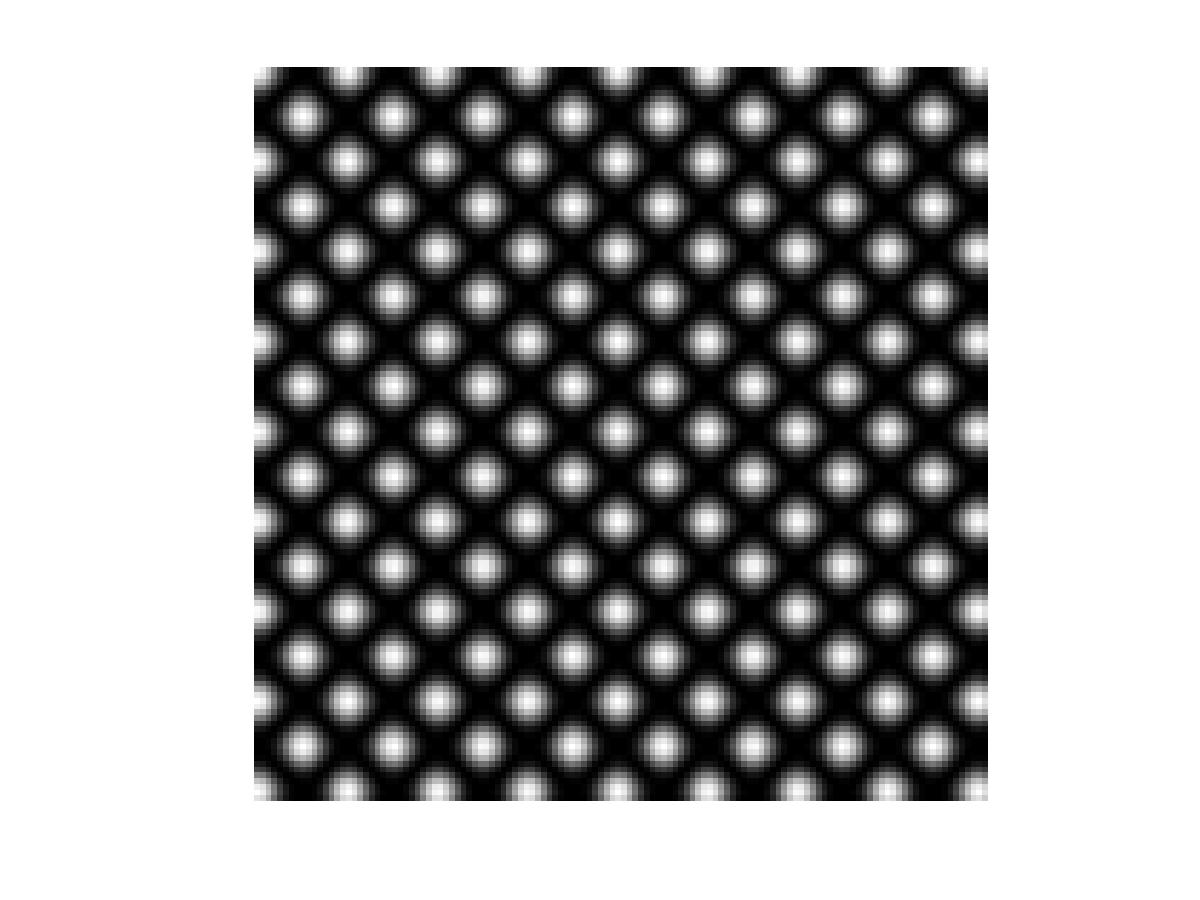}  
 \includegraphics[width=1.6in]{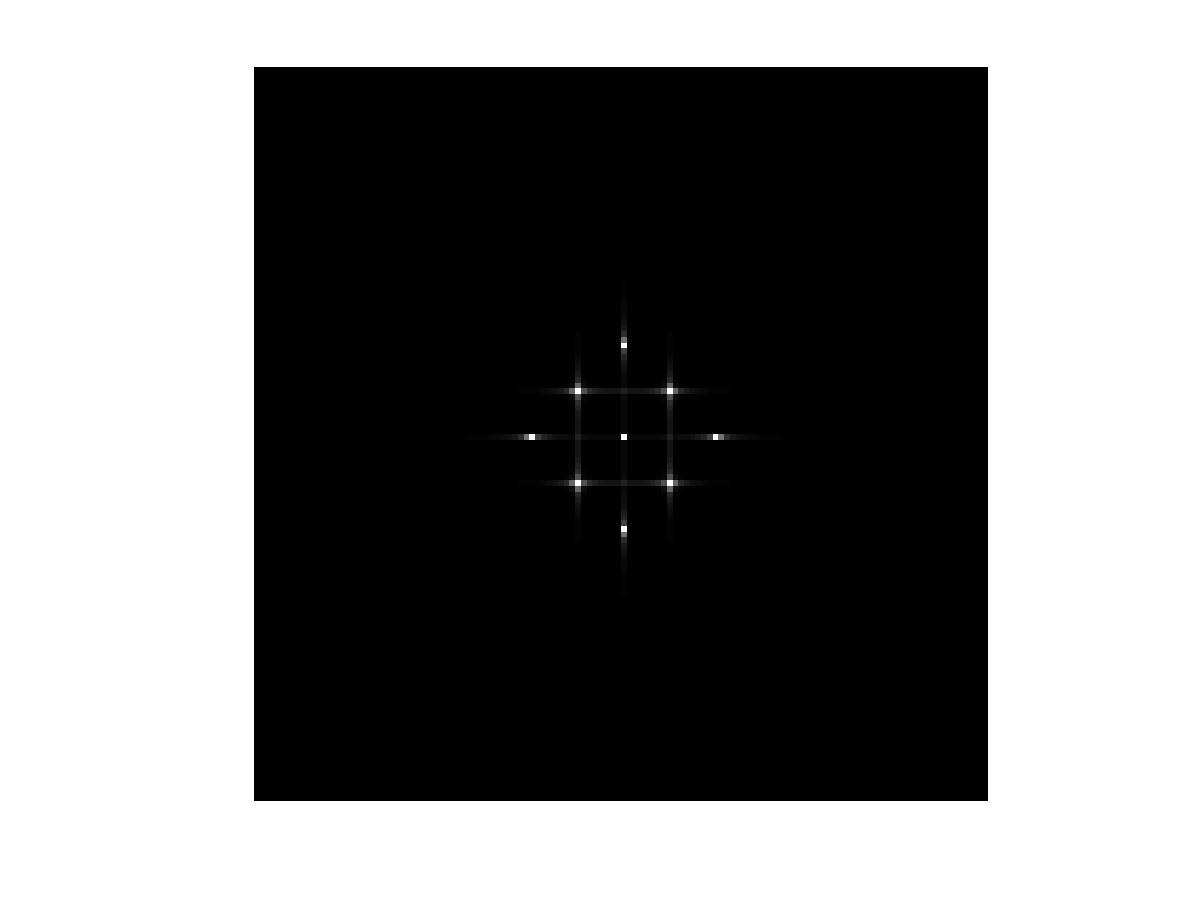} \\[-10pt]
 (a) \hglue 1.5in (b) \hglue 1.4in (c) \hglue 1.5in (d)
 \caption{Image of (a)~first grating; (b)~second grating; (c)~product 
of (a) and (b); (d) { $ \vert {\rm FT} \vert $} of (c).}  \label{fig2}
 \end{figure}

Now let us multiply the amplitude of the gratings and carry out Fourier 
transform on the product signal. It is equivalent to
 \begin{equation}
 {\mathcal FT} [ g_1(x,y)g_2(x,y) ] = {\mathcal FT}[g_1(x,y)] \otimes 
{\mathcal FT}[g_2(x,y)]  \label{eq_prod}
 \end{equation}

\noindent and the simulation results of this operation is displayed 
in Fig.\ref{fig2}.
 The addition signal of Fig.\ref{fig1}(c) and the product signal of 
Fig.\ref{fig2}(c) are different although they are almost similar in 
appearance. That is why the Fourier transform intensity displayed in 
Fig.\ref{fig1}(d) and Fig.\ref{fig2}(d) are different. There are five 
bright dots in Fig.\ref{fig1}(d) whereas there are nine bright dots in 
Fig.\ref{fig2}(d).  The cross terms (bright dots) in Fig.\ref{fig2}(d) 
are due to the convolution between the Fourier transforms of the two 
grating functions.

\subsection{Optical demonstration of convolution}

\begin{figure}
 \centering
 \includegraphics[width=5.5in]{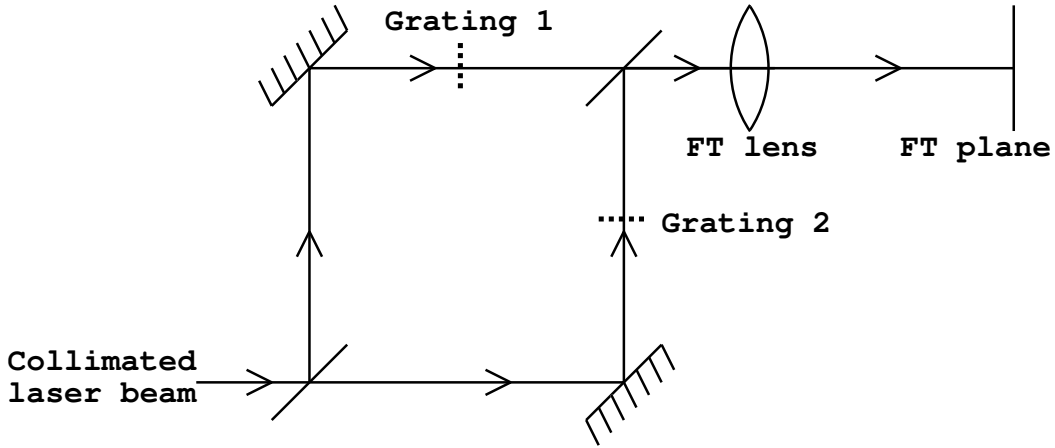}
 \caption{Mach-Zehnder interferometric arrangement for amplitude 
addition and Fourier transform.} \label{MZI}
 \end{figure}

The above examples can be demonstrated using optical experiments. The 
example of equation~(\ref{eq_add}) can be optically implemented using a 
Mach-Zehnder interferometric arrangement as shown in Fig.\ref{MZI}. The 
two grating transparencies are placed inside the two arms of the 
interferometer such that they are equidistant from the second 
beam-splitter. The converging lens acts as a 2D Fourier transforming 
element which produces the Fourier transform of the amplitude distribution 
at the output plane.

\begin{figure}
 \centering
 \includegraphics[width=5.5in]{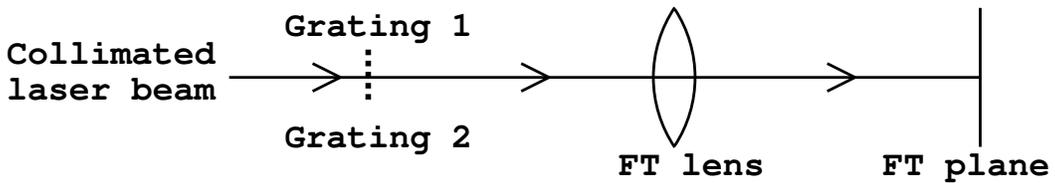}
 \caption{Optical Fourier transform of product of two grating amplitude 
transmittance by a converging lens.} \label{OFT}
 \end{figure}

The convolution theorem of Fourier transform can be optically 
demonstrated using the simple arrangement of optical Fourier 
transformation by a converging lens as shown in Fig.\ref{OFT}.
 The effect of equation~(\ref{eq_prod}) can be yielded by placing the 
grating transparencies in contact with each other at the front focal 
plane of the FT lens in Fig.\ref{OFT}. That is equivalent to multiplying 
the amplitude transmittance functions of the two gratings. Fourier 
transformation by the FT lens produces the convolution of the Fourier 
transforms of the grating transmittance functions at the back focal 
plane which is the output plane.

\subsection{Abbe-Porter experiment and spatial filtering}

The well-known Abbe-Porter experiment is also portrayed as a 
demonstrative experiment of convolution theorem of Fourier 
transform~\cite{Goodman_book}. In this classical experiment, the 
superposition of all the spatial frequency content of an input 2D image 
is recombined after a Fourier transform opreation. The 4f Fourier 
optical set-up with two identical FT lens serves the purpose. The 
 \begin{figure}
 \centering
 \includegraphics[width=5.5in]{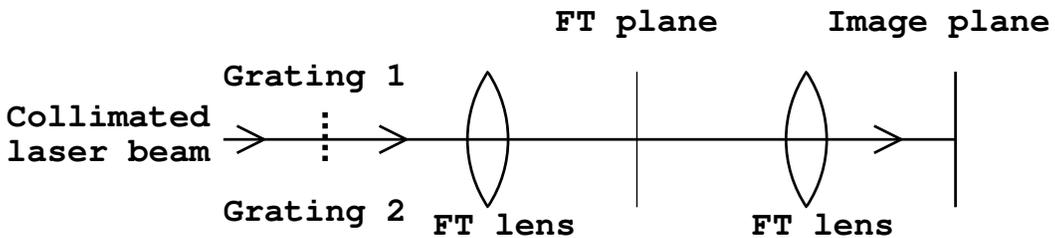}
 \caption{4f Fourier optical set-up to demonstrate Abbe-Porter 
experiment.} \label{4fsetup}
 \end{figure}
 students can play with such a system to understand the role of 
free-space propagation and Fourier transform in imaging. Different slits 
can be placed at the FT plane to study the effect of spatial frequency 
filtering. This can immensely help to improve the understanding of 
convolution theorem of Fourier transform.

\section{2D examples of correlation}

\begin{figure}[htb]
 \centering
 \includegraphics[width=1.6in]{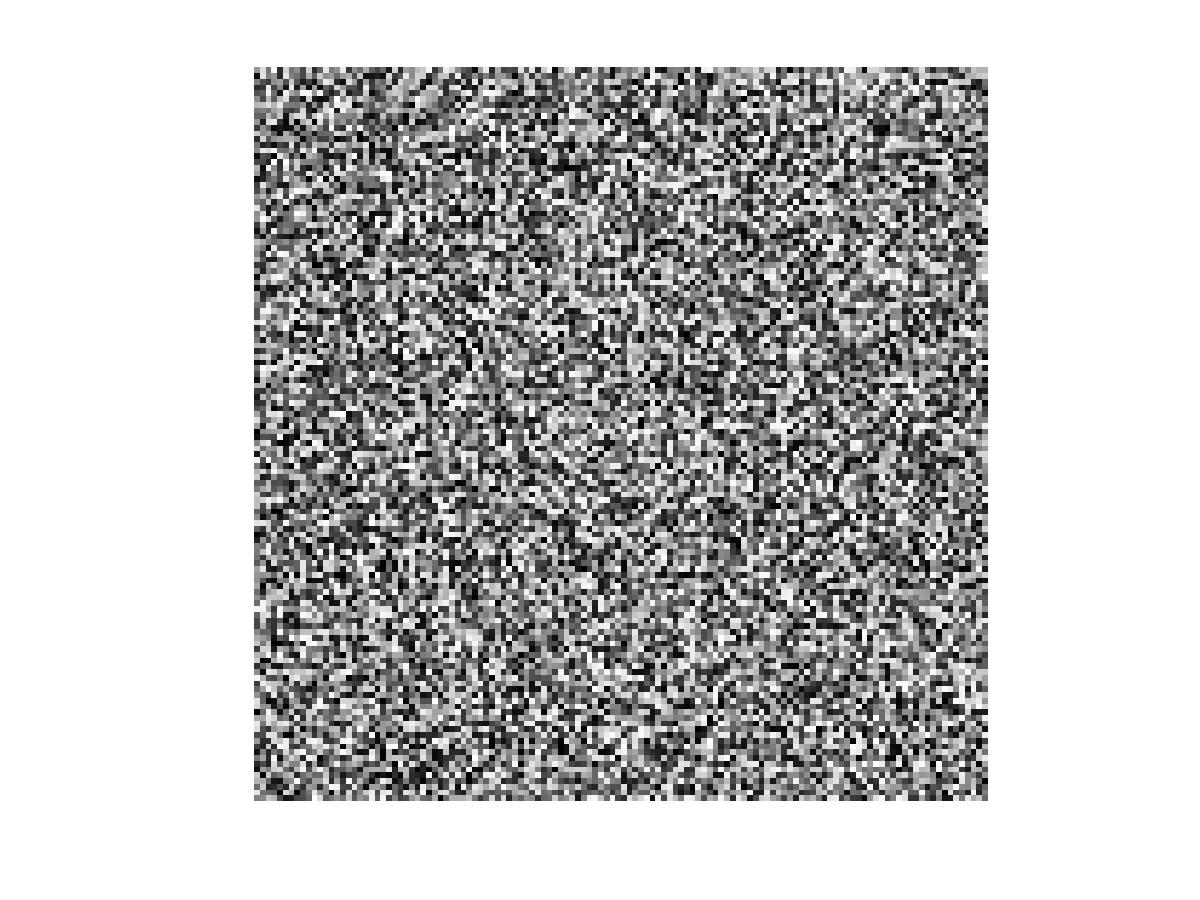}
 \includegraphics[width=1.6in]{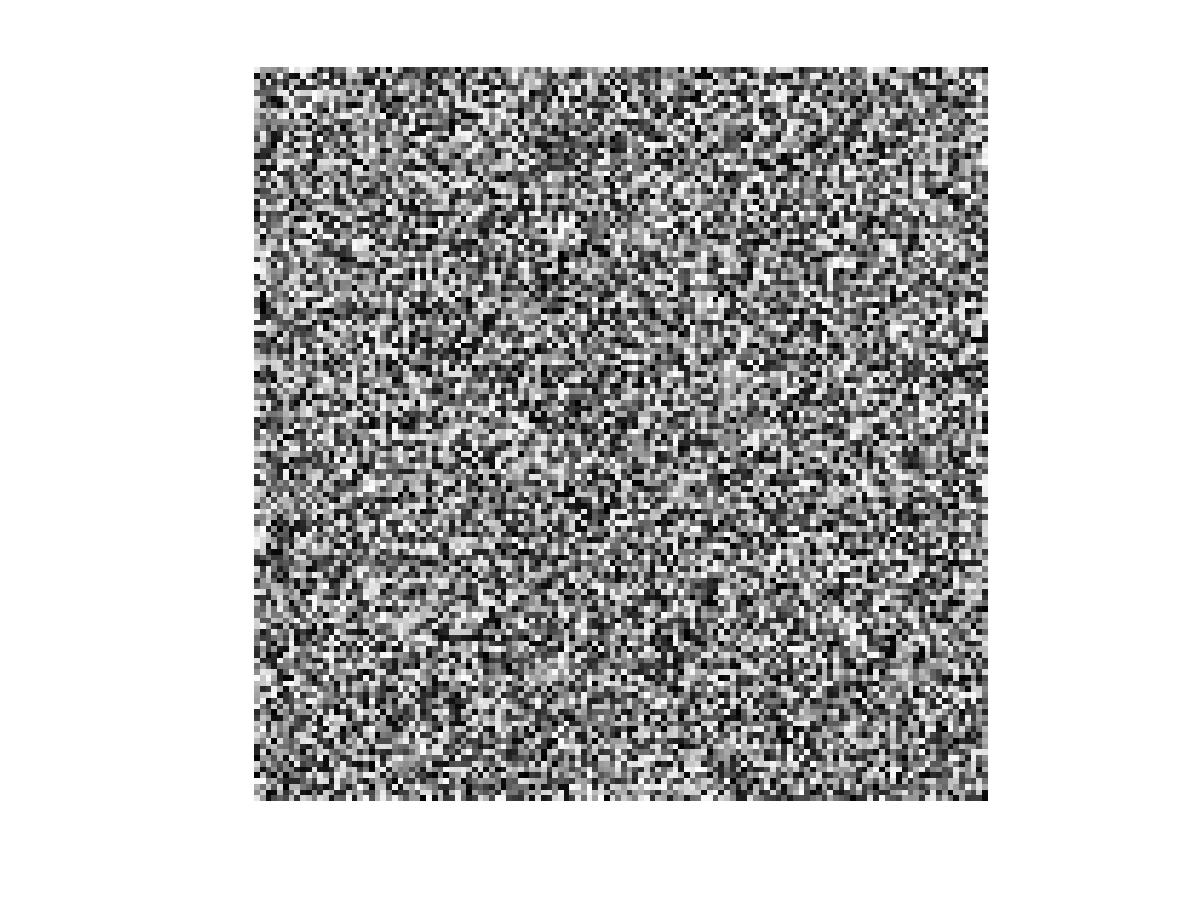}
 \includegraphics[width=1.6in]{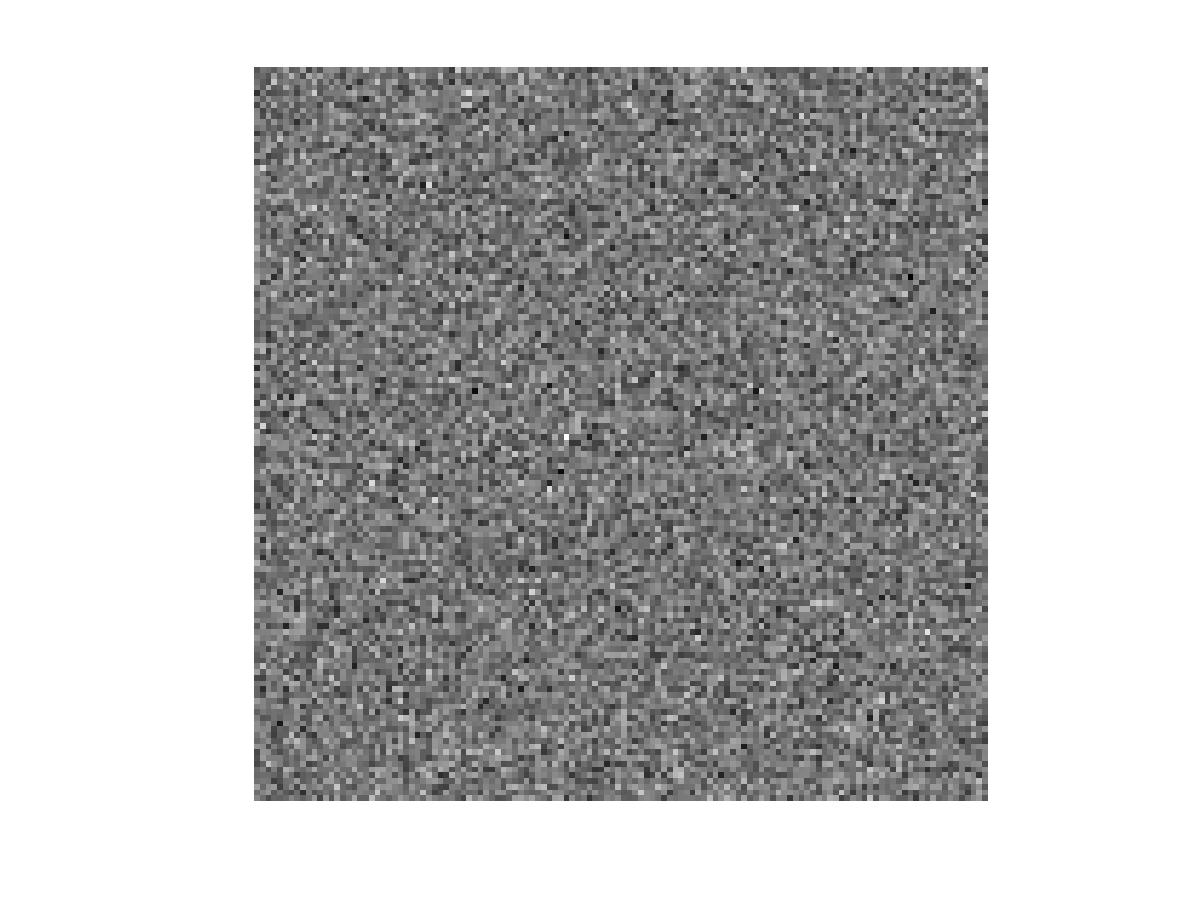} \\[-10pt]
 (a) \hglue 1.5in (b) \hglue 1.4in (c))
 \caption{Image of (a)~first random screen; (b)~second random screen; 
(c)~cross-correlation between (a) and (b).}  \label{randcorfig1}
 \end{figure}

\begin{figure}[htb]
 \centering
 \includegraphics[width=1.6in]{./Pics/gran1.jpg}
 \includegraphics[width=1.6in]{./Pics/gran1.jpg}
 \includegraphics[width=1.6in]{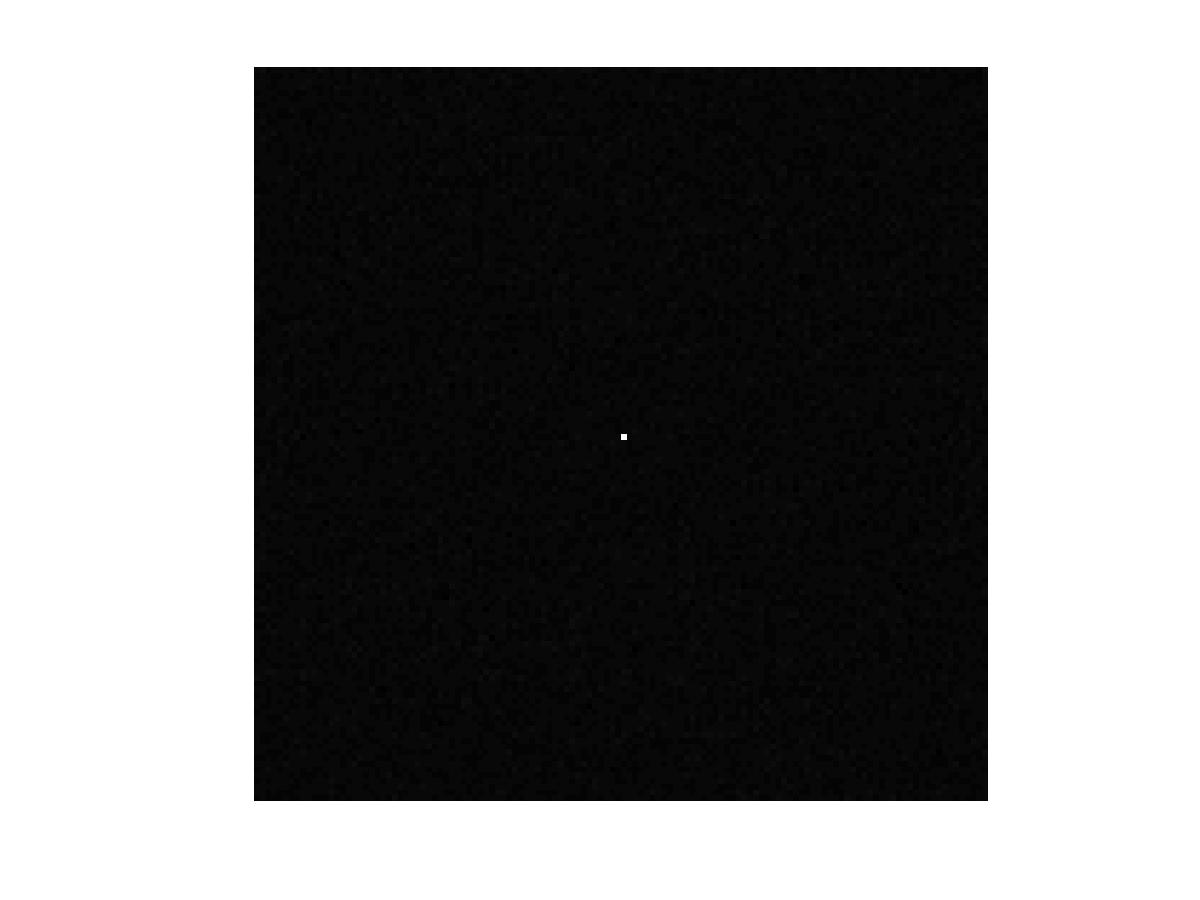} \\[-10pt]
 (a) \hglue 1.5in (b) \hglue 1.4in (c))
 \caption{Image of (a)~first random screen; (b)~identical random screen; 
(c)~cross-correlation between (a) and (b).}  \label{randcorfig2}
 \end{figure}

Correlation operation is handy to measure similarity between two 
signals. Let us take two random screens and inspect the correlation 
between them. We generate images of two random screens using GNU Octave 
as shown in Fig.\ref{randcorfig1}. The cross-correlation between the 
images of two random screens will also produce a random image, because 
the two random images have no correlation. This is shown in 
Fig.\ref{randcorfig1}(c).
 If the two random screens are same, i.e., identical, then the images of 
the random screens are correlated to yield sharp correlation peak. This 
is equaivalent to cross-correlation with itself which is called 
auto-correlation. This result is shown in Fig.\ref{randcorfig2}. The 
presence of a sharp bright dot in Fig.\ref{randcorfig2}(c) is the 
evidence of high correlation between Fig.\ref{randcorfig2}(a) and (b). 

\subsection{Optical correlation} 

The Fourier transforming property of converging lens helps implementing 
correlation operation optically. The Vander Lugt correlation and 
joint-transform correlation can be explained using both simulation and 
experiments. The students can be given mini projects on optical 
correlation.
 \begin{figure}
 \centering
 \includegraphics[width=5.5in]{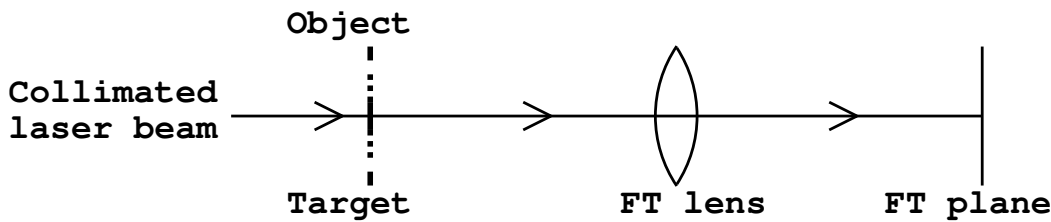}
 \caption{Joint-transform correlation recording set-up.} \label{JTCsetup}
 \end{figure}
  The recording of a joint Fourier transform can be done using a schematic 
set-up similar to Fig.\ref{JTCsetup}. The object and the target optical 
transparencies are placed side by side at the front focal plane of a FT 
lens. The joint Fourier transform intensity is recorded at the FT plane 
 \begin{figure}
 \centering
 \includegraphics[width=5.5in]{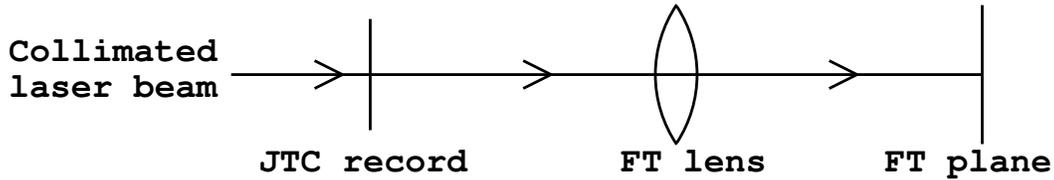}
 \caption{Joint-transform correlation replay set-up.} 
\label{JTCreplay}
 \end{figure}
 either on a photographic plate or on a digital camera sensor. For the 
replay, the photographic record is placed at the front focal plane and 
the cross-correlation output is obtained at the FT plane as shown in 
Fig.\ref{JTCreplay}. If it is a recording on a digital camera, then the 
recorded image of the joint Fourier transform can be displayed on a 
electrically addressed spatial light modulator and joint-transform 
correlation can be obtained by the same set-up..

The Vander Lugt filter is nothing but a Fourier transform hologram of 
the target to be correlated. We can simulate a recording of a Vander 
Lugt filter by recording the Fourier transform of the reference target 
transparency along with a coherent plane reference wave. The Fourier 
transform of the object trapsparency is multiplied with the recorded 
Vander Lught filter and another Fourier transformation produces the 
correlation result. We can see correlation peaks at the location of the 
matched targets. All these exercises can be assigned to the students as 
mini projects which can be carried out using GNU Octave.

\section{Conclusion}

In this paper we have presented a teaching method to explain convolution 
and correlation. It is noticed that convolution can be better explained 
with 2D signals instead of 1D signals. Utilization of interferometric 
system for producing an addition and product of two signals can help to 
demonstrate the convolution theorem of Fourier transform. The difference 
of correlation with convolution can be made clear by demostrating 
cross-correlation operation by Vander Lugt and joint-transform 
correlation. Simulation platform such as GNU Octave can be utilized to 
demonstrate the principles. These exercises can help the students to 
appreciate and understand the concepts of convolution and correlation 
more intuitively.


\begin{thebibliography}{99}

\bibitem{DSP_book_Smith} Smith, S. W., [The Scientist and Engineer's 
Guide to Digital Signal Processing], California Technical Publishing, 
(1999).

\bibitem{VanderLugt} Vander Lugt, A. B., ``Signal detection by complex 
spatial filtering'', IEEE Trans. Info. Theory {\bf IT-10}, 139-145 
(1964).

\bibitem{Goodman_JFT} Weaver, C. J. and Goodman, J. W., ``A technique 
for optically convolving two functions'', Appl. Opt. {\bf 5}, 1248-1249 
(1966).

\bibitem{Goodman_book} Goodman, J. W., [Introduction to Fourier 
Optics, Third Edition], (Roberts \& Company, (2005).

\bibitem{Octave} GNU Octave -- A high-level interactive language for 
numerical computations, http://www.octave.org/

\end{thebibliography}
\end{document}